\documentclass[trackchanges]{aastex701}

\usepackage{amsmath}
\usepackage{multirow}
\usepackage{verbatim}
\usepackage{CJKutf8}
\usepackage{booktabs} 
\usepackage{bm}

\begin{document}

\title{
Data-Constrained Modeling of Electron Transport and Asymmetric Precipitation in the 2011 August 4 Solar Flare
}

\correspondingauthor{Xiangliang Kong}
\email{kongx@sdu.edu.cn}

\author[0000-0002-1576-4033]{Feiyu Yu}
\affiliation{Shandong Key Laboratory of Space Environment and Exploration Technology, Institute of Space Sciences, School of Space Science and Technology, Shandong University, Weihai, Shandong 264209, China}
\email{feiyuyu@mail.sdu.edu.cn}

\author[0000-0003-1034-5857]{Xiangliang Kong}
\affiliation{Shandong Key Laboratory of Space Environment and Exploration Technology, Institute of Space Sciences, School of Space Science and Technology, Shandong University, Weihai, Shandong 264209, China}
\affiliation{Institute of Frontier and Interdisciplinary Science, Shandong University, Qingdao, Shandong 266237, China}
\email{kongx@sdu.edu.cn}

\author[0000-0001-5483-6047]{Ze Zhong}
\affiliation{School of Astronomy and Space Science and Key Laboratory of Modern Astronomy and Astrophysics, Nanjing University, Nanjing 210023, China}
\email{zhong@nju.edu.cn}

\author[0000-0002-4230-2520]{Zhentong Li}
\affiliation{Key Laboratory of Dark Matter and Space Astronomy, Purple Mountain Observatory, Chinese Academy of Sciences, Nanjing 210023, China}
\email{ztli@pmo.ac.cn}

\author[0000-0002-4520-2170]{Zelong Jiang}
\affiliation{Shandong Key Laboratory of Space Environment and Exploration Technology, Institute of Space Sciences, School of Space Science and Technology, Shandong University, Weihai, Shandong 264209, China}
\email{202017715@mail.sdu.edu.cn}

\author[0009-0009-9306-4022]{Yingli Cui}
\affiliation{Shandong Key Laboratory of Space Environment and Exploration Technology, Institute of Space Sciences, School of Space Science and Technology, Shandong University, Weihai, Shandong 264209, China}
\email{202317778@mail.sdu.edu.cn}

\author[0000-0002-6985-9863]{Zhao Wu}
\affiliation{Shandong Key Laboratory of Space Environment and Exploration Technology, Institute of Space Sciences, School of Space Science and Technology, Shandong University, Weihai, Shandong 264209, China}
\email{wuzhao@sdu.edu.cn}

\author[0000-0001-6449-8838]{Yao Chen}
\affiliation{Shandong Key Laboratory of Space Environment and Exploration Technology, Institute of Space Sciences, School of Space Science and Technology, Shandong University, Weihai, Shandong 264209, China}
\affiliation{Institute of Frontier and Interdisciplinary Science, Shandong University, Qingdao, Shandong 266237, China}
\email{yaochen@sdu.edu.cn}

\author[0000-0003-4695-8866]{Gang Li}
\affiliation{State Key Laboratory of Lunar and Planetary Sciences, Macau University of Science and Technology, Macau 999078, China}
\email{gli@must.edu.mo}

\begin{abstract}
Energetic electrons accelerated at coronal reconnection sites during solar flares precipitate into the lower solar atmosphere, generating nonthermal emissions and regulating energy deposition. However, how their transport and precipitation are jointly governed by the three-dimensional (3D) magnetic topology, turbulent scattering, and Coulomb collisions remains unclear. Here, we aim to disentangle these physical processes by using a data-constrained 3D particle transport model for the 2011 August 4 flare. The simulated distribution of precipitated electrons aligns closely with photospheric quasi-separatrix layers and reproduces the observed two-ribbon morphology in 1700~\AA. We reveal a strong polarity asymmetry, with the 10~s precipitation fraction about six times higher in the weak positive polarity. This arises primarily from distinct mirror ratios of different polarities under the 3D magnetic configuration and can be understood via a modified escape probability for an asymmetric magnetic bottle. Varying strengths of turbulent scattering lead to a rise-then-fall trend and a pronounced energy dependence in the precipitation fraction. Coulomb collisions  globally suppress precipitation, especially at low energies, and further amplify the polarity asymmetry. This integrated modeling framework bridges detailed transport physics to observable flare emissions and advances the development of quantitative models for realistic solar flare events.
\end{abstract}


\section{Introduction} \label{sec:intro}
Solar flares are among the most intense energy-release phenomena on the Sun \citep{2011SSRv..159...19F,2017LRSP...14....2B}.
Magnetic reconnection can rapidly convert magnetic energy ($\sim$$10^{32}$~erg over minutes) into plasma heating, nonthermal particle acceleration, and multiwavelength radiation \citep{2017ApJ...836...17A,2020A&A...644A.172W}.
In the standard flare model, particles accelerated near the reconnection region propagate along magnetic loops and precipitate into the lower atmosphere, producing footpoint hard X-ray (HXR) emission and flare ribbons while governing the spatiotemporal distribution of energy deposition \citep{2011SSRv..159..107H}.
The resulting energy input  drives chromospheric evaporation and the formation of hot post-flare loops observable in soft X-rays and extreme ultraviolet (EUV) \citep[e.g.,][]{1994Natur.371..495M, 2013NatPh...9..489S, 2014ApJ...797L..14T}.
Particle transport and precipitation therefore serve as a critical physical link between coronal energy release and multi-wavelength emissions in the lower atmosphere.

Observations frequently show strong asymmetries between conjugate footpoints in intensity, timing, and energy dependence \citep[e.g.,][]{2004A&A...428..219S, 2007A&A...471..705J, 2012ApJ...756...42Y, 2016ApJ...832...63D, 2024SoPh..299..104S}.
These asymmetries may originate from anisotropic injection of energetic electrons, as well as transport effects as electrons travel toward the chromosphere.
Magnetic field convergence near the footpoints narrows the loss cone and traps a fraction of electrons within the loop \citep[e.g.,][]{1998ApJ...505..418F, 2012ApJ...752....4B, 2025FrASS..1110579K}.
Turbulent pitch-angle scattering can deflect electrons into the loss cone, thereby modulating the trapped–to-precipitating partition and the footpoint intensity \citep[e.g.,][]{2014ApJ...780..176K, 2017ApJ...835..262B, 2018ApJ...868L..28E, 2025FrASS..1110579K}.
Meanwhile, Coulomb collisions with the ambient plasma cause energy loss and additional pitch-angle diffusion, effects that are greatly enhanced toward the denser atmosphere and preferentially suppress low-energy electrons \citep[e.g.,][]{2014ApJ...787...86J, 2020ApJ...902...16A, 2022ApJ...941L..22K}.

Flare ribbons typically trace strong gradients in magnetic connectivity, such as quasi-separatrix layers (QSLs; \citealt{1996A&A...308..643D, 2002JGRA..107.1164T,2009ApJ...700..559M, 2015ApJ...810...96S, 2016A&A...591A.141J, 2021NatCo..12.2734Z}).
Footpoint HXR emissions often appear as compact kernels within UV/H$\alpha$ ribbons, although ribbon-structured HXR sources have also been reported in several events \citep[e.g.,][]{2001SoPh..204...55M,2007ApJ...658L.127L,2011ApJ...739...96K}.
Recent studies have increasingly emphasized that realistic 3D magnetic topologies strongly constrain particle transport and emission morphology.
For example, by combining a data-constrained magnetohydrodynamic (MHD) simulation and stereoscopic HXR observations, \citet{2026ApJ...998L..28M} showed that the migration of conjugate HXR footpoint sources along the polarity inversion line observed in the flare implies continuous reconnection along a single QSL system.

Macroscopic models of particle acceleration and transport have recently been developed based on MHD simulations of solar flares. Approaches for particle dynamics coupled with these MHD simulations include solving the transport equations \citep[e.g.,][]{2019ApJ...887L..37K,2020ApJ...905L..16K,2022ApJ...941L..22K,2022ApJ...933...93K, 2022ApJ...932...92L, 2025ApJ...991..202L} or equations of motion under the guiding center approximation \citep[e.g.,][]{2020ApJ...902..147G,2024MNRAS.529.2399B, 2025ApJ...992...81W, 2026A&A...706A..32M}, as well as more self-consistent models incorporating energetic particle feedback onto the background MHD medium \citep[e.g.,][]{2020ApJ...896...97R, 2021PhRvL.126m5101A, 2023SoPh..298..134D, 2024A&A...684A.171D, 2024ApJ...977..146S, 2026ApJ...997..313S, 2026ApJS..283...23H}. 
Accelerating a sufficient population of electrons to high energies likely requires the combined operation of multiple acceleration mechanisms in the current sheet and looptop regions \citep{2025ApJ...991..202L}. In 3D configurations, the ubiquitous occurrence of magnetic turbulence and instabilities \citep[e.g.,][]{2022NatAs...6..317S, 2023ApJ...954L..36W} further complicates particle transport dynamics, leading to the spatiotemporal complexity of HXR emission and flare ribbons.

In this Letter, we investigate the propagation and precipitation of energetic electrons during the 2011 August 4 flare by combining the focused transport equation (FTE) with a static snapshot adopted from a data-driven 3D MHD model.
We compare the simulated electron precipitation distribution with the photospheric QSLs and the 1700~\AA\ flare ribbons observed by the Atmospheric Imaging Assembly (AIA; \citealp{2012SoPh..275...17L}).
Using temporal evolution and quantitative analysis, we aim to disentangle the roles of different physical processes in particle transport and precipitation.
The paper is organized as follows. We describe the numerical methods in Section~\ref{sec:methods}, and then present our simulation results in Section~\ref{sec:results}. Finally, conclusions and discussion are given in Section~\ref{sec:conclusion}.

\section{Numerical Methods} \label{sec:methods}

\begin{figure}[!htbp]
\centering
\includegraphics[width=0.95\textwidth]{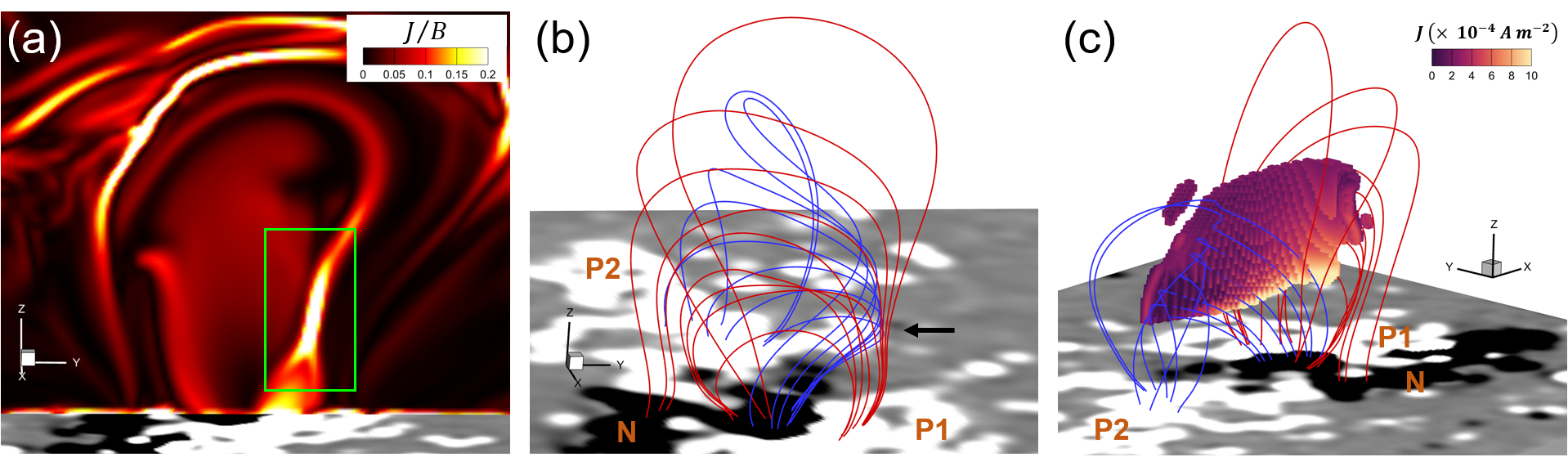}
\caption{3D magnetic configuration and current sheet at 03:52~UT in the MHD model.
(a) Distribution of $J/B$ in a representative $y$–$z$ slice; the green box highlights the reconnection current sheet. 
(b) Representative magnetic field lines overlaid on the bottom $B_z$ map, illustrating conjugate connectivity linking positive-polarity footpoints $P1$ and $P2$ to the negative-polarity footpoint $N$; the black arrow points to the X-type geometry at the reconnection site. (c) Extracted 3D current sheet structure colored by $J$, showing stronger current density toward lower heights.
}
\label{fig1}
\end{figure}

\begin{figure}[!htbp]
\centering
\includegraphics[width=0.95 \textwidth]{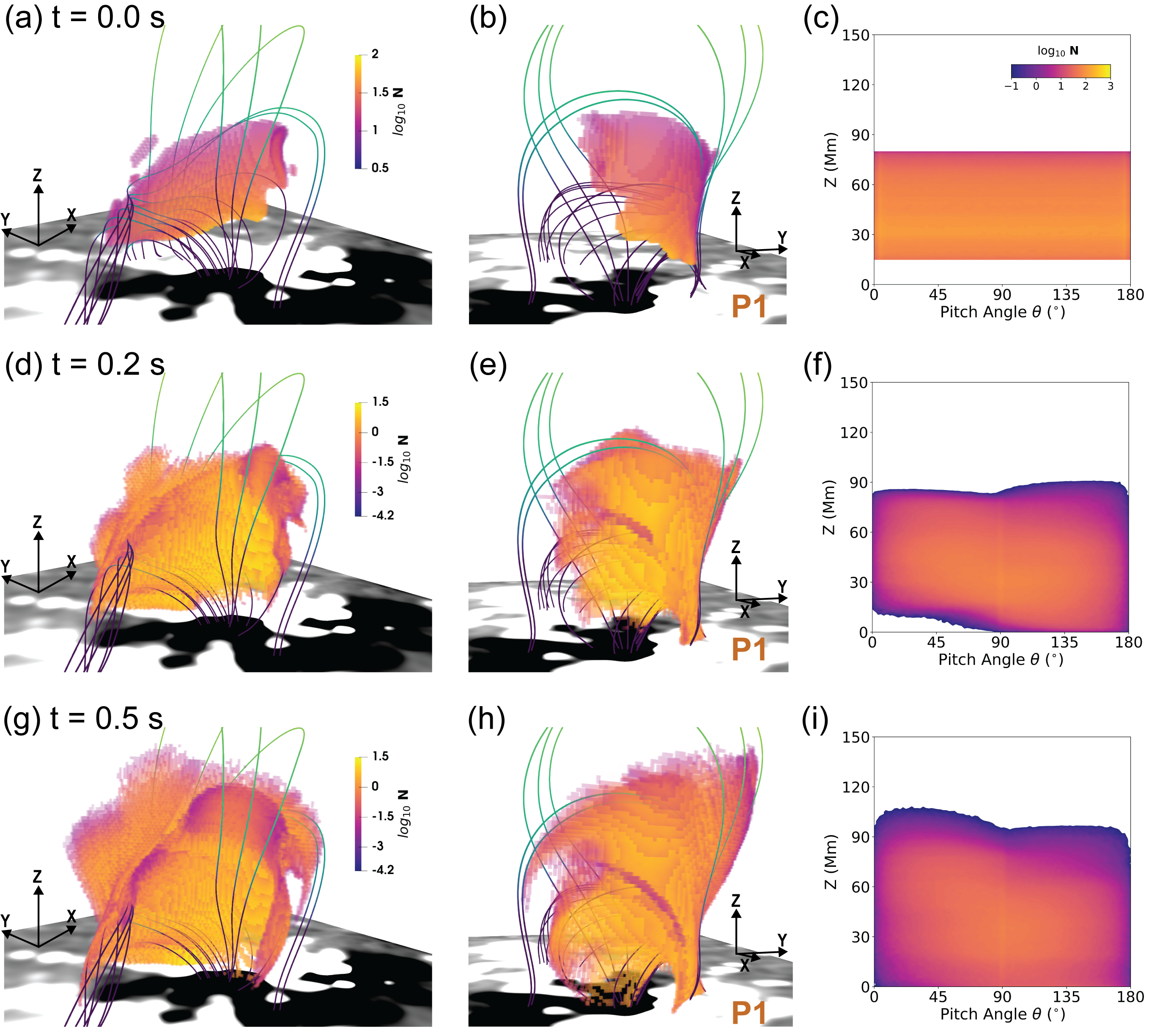}
\caption{
Early transport of 20–50~keV electrons in the 3D magnetic field at $t=$ 0, 0.2, and 0.5~s post-injection. Left and middle columns display the spatial distributions from two viewing angles. Right column presents the horizontally integrated distribution plotted against height $z$ and pitch angle $\theta$. 
}
\label{fig2}
\end{figure}

\begin{figure}[!htbp]
\centering
\includegraphics[width=0.95\textwidth]{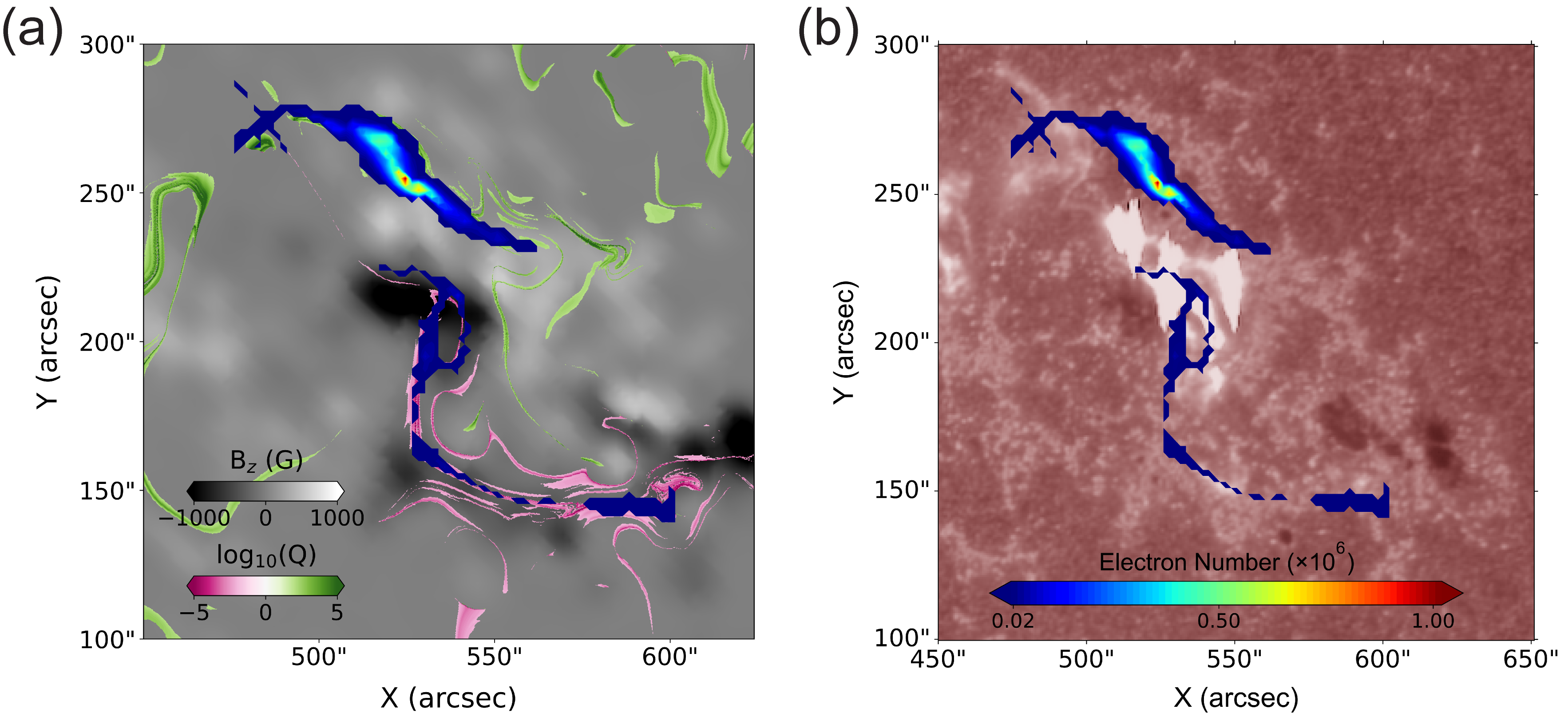}
\caption{
Distribution of 10~s accumulated precipitating electrons mapped to the magnetogram at the bottom boundary and compared with the photospheric QSLs (a), and overlaid on the observed AIA 1700~\AA\ image at 03:52~UT (b). 
}
\label{fig3}
\end{figure}

\begin{figure}[!htbp]
\centering
\includegraphics[width= 0.85\textwidth]{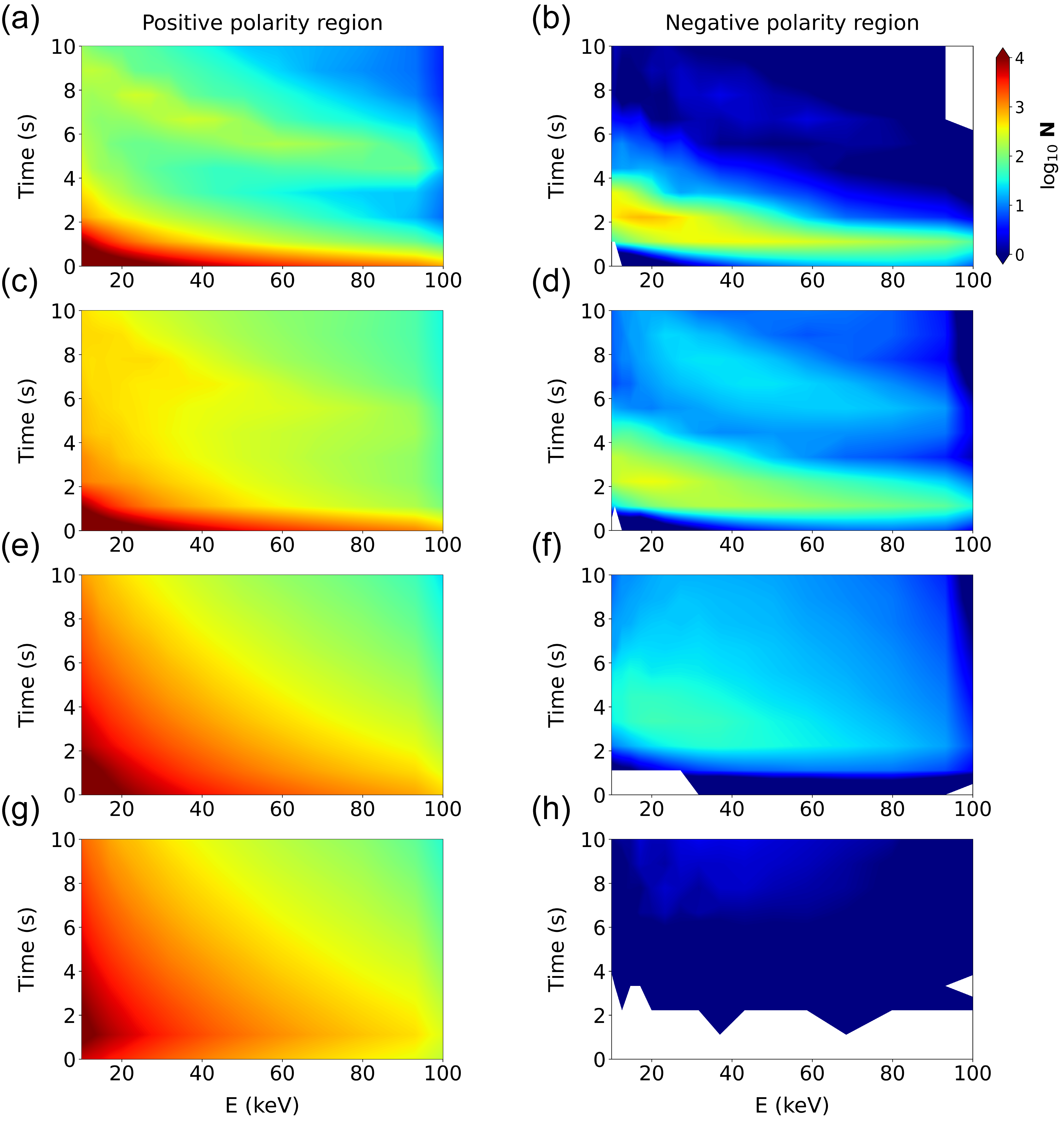}
\caption{
Temporal evolution of precipitating electron counts at different energies under four turbulent scattering regimes: (a)-(b) no scattering ($D_{\mu\mu0}^{t}=0$~s$^{-1}$), (c)-(d) weak scattering ($D_{\mu\mu0}^{t}=0.01$~s$^{-1}$), (e)-(f) moderate scattering ($D_{\mu\mu0}^{t}=0.5$~s$^{-1}$), and (g)-(h) strong scattering ($D_{\mu\mu0}^{t}=5$~s$^{-1}$). Left and right columns are for the positive and negative polarities, respectively. 
}
\label{fig4}
\end{figure}

\begin{figure}[!htbp]
\centering
\includegraphics[width=0.95\textwidth]{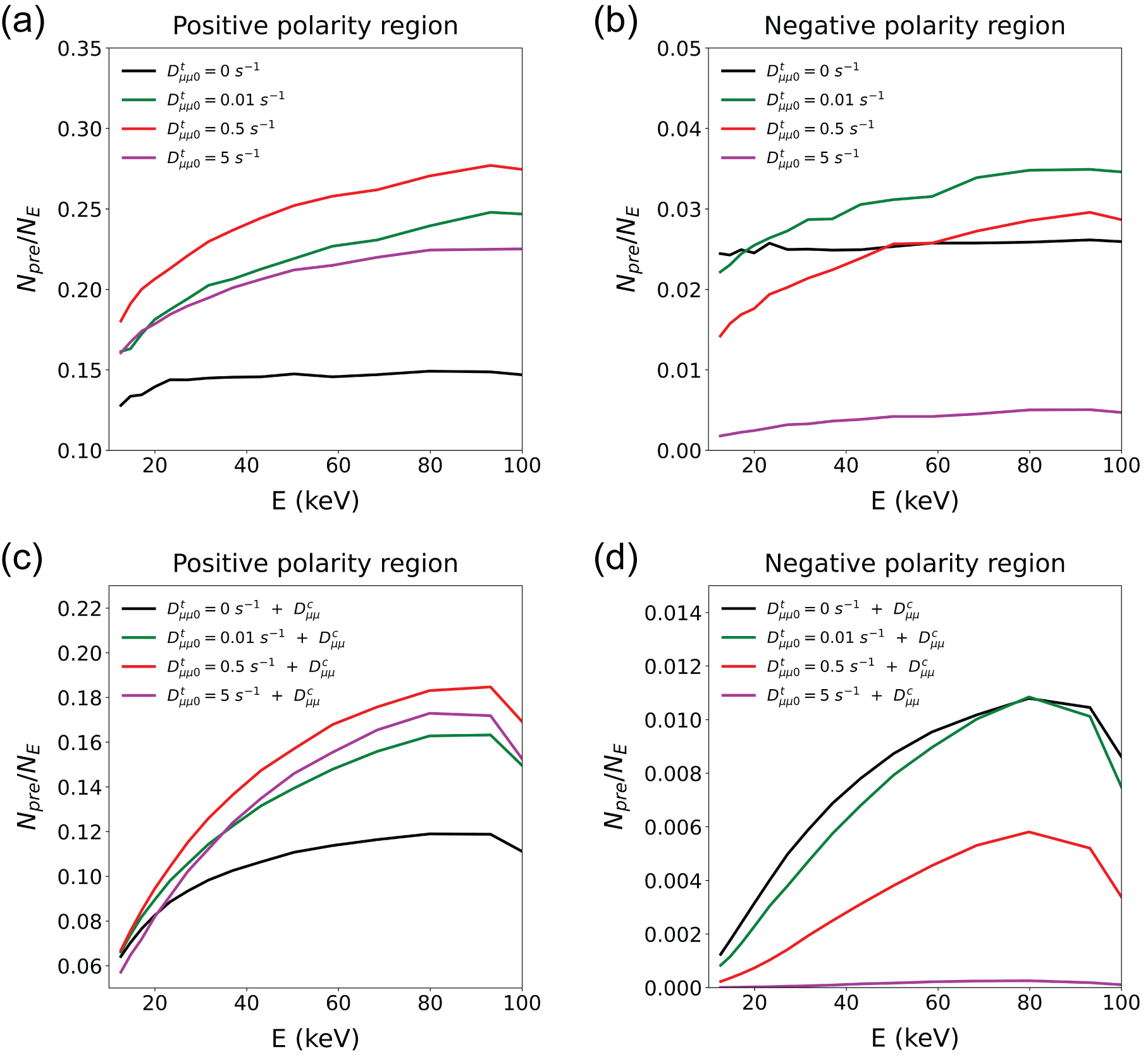}
\caption{
Energy dependence of the 10~s accumulated electron precipitation fraction $N_{\rm pre}/N_E$ in positive and negative polarities under different scattering regimes ($D_{\mu\mu0}^{t}=0$, 0.01, 0.5, and 5~s$^{-1}$), with Coulomb collisions neglected in upper panels (a)-(b) and included in lower panels (c)-(d).
}
\label{fig5}
\end{figure}

We perform a data-driven 3D MHD simulation of the 2011 August 4 flare using version 2.0 of the Message Passing Interface Adaptive Mesh Refinement Versatile Advection Code (MPI-AMRVAC; \citealp{2018ApJS..234...30X}) under the zero-$\beta$ approximation \citep{2019ApJ...870L..21G, 2021ApJ...919...39G, 2021NatCo..12.2734Z} to obtain the evolution of coronal magnetic field and plasma.
The data-driven setup and numerical implementation follow \cite{2023ApJ...947L...2Z}. The simulation uses a uniform grid at 2$\arcsec$ resolution and is initialized with a nonlinear force-free field extrapolation. The bottom boundary is driven by a time series of Helioseismic and Magnetic Imager (HMI; \citealp{2012SoPh..275..207S}) vector magnetograms taken at a 720 s cadence.

The M9.3 flare started at 03:42 UT and peaked at 03:58 UT. We use the snapshot at 03:52~UT during the impulsive phase, when the current sheet and flare loops are well developed, as the background for particle transport modeling.
The computational domain covers both the core active region and the overlying corona.
In addition to the 3D magnetic fields, the MHD data provide the density stratification required for Coulomb collisions. Because tens-of-keV electrons travel much faster than the Alfvén speed, their transit time through the flare loop is far shorter than the typical MHD timescale. We therefore treat the MHD background as static and focus on the early transport and precipitation of promptly injected electrons within a 10~s time interval. We refer to this approach as a data-constrained 3D particle transport model.

Figure~\ref{fig1} displays the 3D magnetic configuration and current sheet at 03:52~UT.
We identify the reconnection current sheet using the normalized ratio of current density to magnetic field strength, $J/B$ \citep{2016ApJ...828...62J, 2023ApJ...947L...2Z, 2023ApJS..266....3G}. 
This ratio effectively captures regions of localized magnetic shear and steep gradients within current layers in 3D models and can be expressed as $(\Delta B/B)/\Delta x$, where $\Delta B$ is the magnetic field change between adjacent grids, and $\Delta x$ is the local grid spacing.
The reconnection and outflow regions are favorable sites for electron acceleration via multiple mechanisms \citep[e.g.,][]{1996ApJ...462..997L, 2019ApJ...887L..37K, 2025ApJ...991..202L}.
We define the particle injection volume as the reconnecting current sheet beneath the flux rope, as shown in Figure~\ref{fig1}(c).

We use the median to characterize magnetic field strength in different regions, thereby minimizing the influence of outliers. The median magnetic field strength in the current sheet is $B_{\rm cs} \approx 31$~G.
At the bottom boundary, the median magnetic fields are $B_p \approx 218$~G in the positive polarity and $B_n \approx -315$~G in the negative polarity. This indicates distinct mirror ratios of the two polarities, which provide a natural geometry for asymmetric electron precipitation.

In the corona, the gyroradii of nonthermal electrons ($10^{-1}-10^{-2}$~m) are much smaller than the macroscopic scale of flares ($\sim$$10^8$~m). Following our previous work  \citep{2022ApJ...941L..22K}, the evolution of the electron distribution function $f(\mathbf{X},\mu,p,t)$ is described on the prescribed MHD background using the FTE, where $\mathbf{X}$ is position, $\mu$ is the pitch-angle cosine, and $p$ is momentum.
The transport equation is written as \citep[e.g.,][]{1971ApJ...170..265S,2006JGRA..111.8101Q},
\begin{equation}
\frac{\partial f}{\partial t}
= -\left(v\mu\hat{\mathbf{b}}+\mathbf{U}\right)\cdot\nabla f
+ \frac{\partial}{\partial\mu}D_{\mu\mu}\frac{\partial f}{\partial\mu}
- \frac{d\mu}{dt}\frac{\partial f}{\partial\mu}
- \frac{dp}{dt}\frac{\partial f}{\partial p}.
\end{equation}
where $\hat{\mathbf{b}}$ denotes the direction of the unit magnetic field and $\mathbf{U}$ is the flow velocity.
Magnetic mirroring enters through the term $d\mu/dt$ and the momentum change term $dp/dt$ includes adiabatic effects and Coulomb energy losses.

The pitch-angle diffusion coefficient contains two components, ${D_{\mu \mu}} = D_{\mu \mu}^t + D_{\mu \mu}^c$.
$D_{\mu \mu}^t$ represents resonant scattering by turbulent magnetic fields in quasi-linear theory \citep{1971RvGSP...9...27J}. In the nonrelativistic limit, the turbulent diffusion is written as \citep{1986ApJ...311..437B, 2022ApJ...941L..22K}:
\begin{equation}
D_{\mu \mu}^t = D_{\mu \mu 0}^t (1 - \mu^2)(|\mu|^{\Gamma -1} + h_0) \left(\frac{B}{B_0} \right)^{2 - \Gamma} \left(\frac{p}{p_0} \right)^{\Gamma -1},
\end{equation}
where $D_{\mu\mu0}^{t}$ sets the magnitude of scattering for 5 keV electrons (with momentum $p_0$), $\Gamma$ = 5/3 is the spectral index of the Kolmogorov turbulence spectrum, and $h_0$ = 0.05 ensures finite scattering near $\mu \approx 0$.

When electrons propagate into the dense solar atmosphere, Coulomb collisions can modify both their pitch angles and energies. For fully ionized plasma, the collisional pitch-angle diffusion coefficient $D_{\mu \mu}^c$ is given by \citep{2014ApJ...780..176K}:
\begin{equation}
D_{\mu\mu}^{c}=\frac{2K\,n_{\rm th}}{m_e^{2}v^{3}}(1-\mu^{2}),
\end{equation}
where $K = 2\pi e^{4} \Lambda$, $\Lambda$ is the Coulomb logarithm, and $n_{\rm th}$ is the thermal density.
The nonrelativistic Coulomb energy loss rate is \citep{1971SoPh...18..489B, 1978ApJ...224..241E}:
\begin{equation}
\frac{dE}{dt}= -\frac{K}{E}\,n_{\rm th}\,v,
\end{equation}
where $E$ is the electron energy and in the nonrelativistic limit, $E = p^2/2m_e$.
Both collisional diffusion and energy loss increase rapidly toward the dense lower atmosphere and preferentially affect low-energy electrons.

To classify scattering regimes, we compare a characteristic scattering time $\tau_d=1/D_{\mu\mu}$ with a loop-crossing time $\tau_c$ \citep{1987SoPh..114..127B}.
For consistency across parameter sets, $\tau_d$ is evaluated by a representative $D_{\mu\mu}$ near the current sheet and loop apex. 
We adopt a characteristic loop length $L=80$~Mm.
For 20~keV electrons, $v_{e} \approx 0.272\ c$, giving $\tau_c =L/v_{e} \approx 0.98$~s.
We refer to weak scattering when $\tau_d>\eta\,\tau_c$, moderate scattering when $\tau_c<\tau_d<\eta\,\tau_c$, and strong scattering when $\tau_d<\tau_c$, where $\eta$ is the magnetic mirror ratio.
We consider four turbulent scattering levels: $D_{\mu\mu0}^{t} = 0,\ 0.01,\ 0.5$, and 5~s$^{-1}$, covering weak to strong scattering regimes.

The FTE is solved numerically by integrating pseudo‑particle trajectories in the 3D MHD magnetic fields using time-forward stochastic differential equations \citep[e.g.,][]{1999ApJ...513..409Z, 2017SSRv..212..151S, 2022ApJ...941L..22K}, which is implemented as an independent \emph{Fortran} post-processing tool.
We do not explicitly model electron acceleration, but instead prescribe a nonthermal source via injection into the extracted current sheet volume (Figure~\ref{fig1}(c)) and focus on subsequent electron transport and precipitation processes.
For each simulation, $4\times10^{6}$ pseudo-electrons with a power-law spectrum $f(E)\propto E^{-2}$ ranging from 5 to 100~keV are isotropically injected. We then track the 3D spatial distribution and precipitation location of electrons, and separately count the number of electrons precipitating into the positive and negative polarities over a 10 s time window, from which the precipitation fraction is calculated.

\section{Simulation Results} \label{sec:results}

We first investigate the early evolution of injected electrons in the 3D magnetic field under the moderate diffusion regime ($D_{\mu\mu0}^{t}=0.5$~s$^{-1}$) with Coulomb collisions.
Figure~\ref{fig2} shows the temporal evolution of the 20–50~keV electron distribution and the corresponding count map in $z$–$\theta$ space.
Immediately after injection (Figures~\ref{fig2}(a)-(b)), the electrons remain confined within the current sheet, exhibiting an isotropic pitch-angle distribution (Figure~\ref{fig2}(c)).
By $t=0.2$~s (Figures~\ref{fig2}(d)-(e)), the first precipitation of electrons occurs in the positive-polarity field ($P1$). In the $z$–$\theta$ map (Figure~\ref{fig2}(f)), most electrons at lower heights ($<$10 Mm) are characterized by $\theta > 90^{\circ}$. 
In contrast, electron counts in the negative-polarity field ($N$) show a notable increase only at $t=0.5$~s, when precipitated electrons at the bottom boundary exhibit pitch angles $<90^{\circ}$ (Figures~\ref{fig2}(g)-(i)).
Note that these early precipitated electrons come mainly from the lowest altitude of the current sheet. Their polarity-dependent onset occurs under isotropic particle injection and arises from combined effects including the distinct magnetic field strength of the two polarities and the asymmetric loop geometry.

Figure~\ref{fig3} compares the distribution of precipitated electrons accumulated over 10~s with the photospheric QSLs and the AIA 1700~\AA\ flare ribbons.
The QSLs are derived from the 3D MHD magnetic field using the squashing factor $Q$. Regions satisfying $|\log(Q)|>3$ are overplotted to highlight strong gradients in magnetic connectivity \citep{2023ApJ...947L...2Z}, which are likely linked to the reconnection current sheet.
As shown in Figure~\ref{fig3}(a), the precipitating electrons form elongated ribbon-like structures that align well with the photospheric QSLs. This indicates that electrons accelerated near the current sheet propagate sunward along newly reconnected magnetic field lines and deposit their energy in the low atmosphere.
To directly compare the simulation results with observations, we project the AIA 1700~\AA\ image observed at 03:52~UT onto the bottom boundary of the MHD model using the coordinate mapping procedure detailed in \citet{2023ApJ...947L...2Z}, thus achieving a pixel-to-pixel correspondence.
As shown in Figure~\ref{fig3}(b), the simulation  successfully reproduces both the overall two-ribbon morphology and the fragmented fine structures.
The negative polarity ribbon matches the observation well, whereas the positive polarity ribbon exhibits a small spatial offset.
These results demonstrate that the spatial distribution of electron precipitation is controlled by the 3D magnetic connectivity, which bridges coronal magnetic reconnection to the formation of chromospheric flare ribbons.

Figure~\ref{fig4} shows the temporal evolution of electron counts precipitated in positive and negative polarities for various turbulent scattering coefficients. 
Without turbulent scattering (Figures~\ref{fig4}(a)-(b)), a distinct polarity asymmetry emerges between the positive (left) and negative (right) polarity maps.
Electrons precipitate earlier and more efficiently in the positive polarity, whereas precipitation in the negative polarity is delayed and significantly suppressed. This is mainly due to stronger magnetic trapping when electrons stream toward the more intense negative polarity. 
Meanwhile, Coulomb collisions primarily impact the energies of low-energy electrons, leading to greater suppression of low-energy precipitation.
In the weak scattering regime ($D_{\mu\mu0}^{t}=0.01$~s$^{-1}$, Figures~\ref{fig4}(c)-(d)), the electron counts increase in both polarities relative to the non-scattering case. Owing to turbulent scattering and collisional diffusion, magnetically trapped electrons are continuously scattered into the loss cone, thereby gradually precipitating into both polarities.
Under moderate scattering ($D_{\mu\mu0}^{t}=0.5$~s$^{-1}$, Figures~\ref{fig4}(e)-(f)), electrons are resupplied to the loss cone in a more steady and persistent manner, producing a smoother temporal evolution of electron counts. Nevertheless, the negative polarity exhibits suppressed precipitation at early times, with the low-energy component being nearly negligible.
When scattering is further strengthened ($D_{\mu\mu0}^{t}=5$~s$^{-1}$, Figures~\ref{fig4}(g)-(h)), the rapid variation of electron pitch angles strongly hinders the net downward transport. Early electron precipitation is suppressed in both polarities, resulting in the complete absence of precipitating electrons across all energies in the negative polarity.

We further quantify electron precipitation by defining the precipitation fraction as the ratio of the 10~s accumulated electron counts in positive and negative polarities to the total number of injected electrons, $N_{\rm pre}/N_E$.
To separate the effects of turbulent scattering and Coulomb collisions, we perform simulations without and with Coulomb collisions, as shown in the upper and lower panels in Figure~\ref{fig5}.

In the absence of turbulent scattering and Coulomb collisions, the electron precipitation fraction is mainly determined by the mirror ratios of the two polarities. In our simulations (Figures~\ref{fig5}(a)-(b), black curves), it reaches $\sim14\%$ in the weaker field (positive polarity), but only $\sim2.5\%$ in the stronger field (negative polarity). 
Theoretically, for an idealized symmetric magnetic mirror system (magnetic bottle), the escape probability of isotropically distributed electrons from one side of the loss cones is
\begin{equation}
P(\eta)=\frac{1}{2}\left(1-\sqrt{1-\frac{1}{\eta}}\right),
\end{equation}
where $\eta$ is the magnetic mirror ratio, i.e., the magnetic field strength at the footpoints divided by that in the current sheet.

As shown in Figure~\ref{fig3}(a), the electron precipitation locations closely follow the photospheric QSLs. The median magnetic field strength within these regions is $B_p \approx 110$~G in the positive polarity and $B_n \approx -318$~G in the negative polarity.
The corresponding mirror ratios are then $\eta_p=B_p/B_{\rm cs}\approx 3.5$ and $\eta_n=|B_n|/B_{\rm cs}\approx 10.3$, yielding escape probabilities of $P_p \approx7.7\%$ and $P_n \approx2.5\%$, respectively.
Thus, the precipitation fraction is consistent with theoretical predictions in the negative polarity, but is much higher in the positive polarity.
Due to the asymmetry of the magnetic bottle, particles reflected by the strong negative magnetic field can escape through the weak positive magnetic field (with a fraction of $P_p- P_n$), whereas the converse does not hold.
The modified escape probability in the positive polarity becomes $P_p^{'} = 2 P_p - P_n = 12.9 \%$. 
In addition, the reconnection current sheet where particles are initially injected lies closer to the positive polarity ($P1$), which may also cause more particles to precipitate toward it.

When turbulent pitch-angle scattering is present, the electron precipitation fraction shows a marked energy dependence, increasing with electron energy. This behavior can be understood in terms of the particle mean free path, $\lambda_{\parallel}$ = $3\kappa_{\parallel} / v \propto v^{1/3}$, where $\kappa_{\parallel}$ is the parallel diffusion coefficient and related to $D_{\mu\mu}^{t}$ \citep{1966ApJ...146..480J, 2022ApJ...941L..22K, 2022ApJ...925L..13Y}. Thus, high-energy electrons have a larger mean free path and precipitate more efficiently onto the solar surface.
Furthermore, the precipitation fraction exhibits a “rise-then-fall” trend with increasing scattering strength. 
As shown in Figure~\ref{fig5}(a), for the positive polarity, the precipitation fraction in the weak scattering regime increases by a factor of $\sim1.2$–1.7 over the entire energy range. Under moderate scattering, the positive-polarity fraction attains its maximum value among all cases, reaching nearly 30\% at high energies. 
In contrast, for the negative polarity (Figure~\ref{fig5}(b)), the increase is relatively modest and the value even falls below that in the scattering-free case under moderate and strong scattering, particularly at low energies.
This indicates that the negative-polarity region is more sensitive to turbulent scattering, mainly due to the larger magnetic mirror ratio (i.e., smaller loss cone, critical pitch angle $\theta_c < 20^{\circ}$) and longer loop length (i.e., longer transit time), which makes it easier to satisfy the strong scattering condition and reduces the downward transport of electrons.

Figures~\ref{fig5}(c)-(d) present the simulation results including the effects of Coulomb collisions (i.e., energy losses and collisional pitch-angle scattering). Collisions greatly increase the effective scattering rate, with their impact strengthening as electrons penetrate deeper into the atmosphere. As a result, the precipitation fraction decreases substantially in all simulations. 
Due to greater energy losses and stronger collisional scattering experienced by low-energy electrons, the reduction is more significant at lower energies. For example, the difference in the precipitation fraction between high- and low-energy electrons in the positive polarity increases from a factor of $\sim$1.6 to $\sim$3.
While the overall trend in the positive polarity is nearly unchanged, the fraction in the negative polarity involving turbulent scattering is consistently lower than that without turbulent scattering. 
This effect also amplifies the difference in precipitation between the two polarities, with the ratio of their maximum precipitation fractions increasing from $\sim$8 to $\sim$16.

\section{Conclusions and Discussion} \label{sec:conclusion}
We have established a modeling framework using particle transport solutions constrained by data-driven 3D MHD background fields. This framework can simulate electron propagation and precipitation in realistic solar flare events and facilitate direct model-observation comparisons.
Application to the 2011 August 4 flare shows that the distribution of precipitated electrons closely traces the photospheric QSLs and successfully reproduces the two-ribbon pattern observed in AIA 1700~\AA.

Under the influence of the 3D magnetic field configuration, electron transport and precipitation differ substantially between the two polarities. In the weaker positive polarity, electrons precipitate earlier and in greater numbers.
We calculate the precipitation fraction using the cumulative number of precipitating electrons over 10 s, and find that this fraction in the positive polarity is a factor of 6 larger than that in the negative polarity. The value in the negative polarity is consistent with the escape probability predicted by standard magnetic mirror theory, whereas that in the positive polarity greatly exceeds the theoretical prediction. For this asymmetric magnetic bottle, we modify the escape probability on the weak-field side as $P_p^{'} = 2 P_p - P_n$, which gives results that generally agree with our simulations.

With varying turbulent scattering strengths, the precipitation fraction shows a “rise-then-fall” trend and a clear energy dependence.
For the positive polarity, the precipitation fraction is highest under moderate scattering and approaches nearly 30\% at high energies.
However, the behavior differs somewhat in the negative polarity.
This is because the strong magnetic field results in a small loss cone, and the longer transit time further facilitates the transition to strong scattering regimes, making it difficult for electrons to enter the loss cone and precipitate.
Coulomb collisions lead to energy loss and collisional scattering, lowering the overall precipitation fraction with stronger suppression at lower energies. In particular, for the negative polarity, once Coulomb collisions are taken into account, the presence of turbulent scattering reduces the precipitation fraction in all cases.

We adopt a set of diffusion coefficients to represent various transport regimes covering non-scattering, weak, moderate, and strong turbulent scattering conditions \citep{1987SoPh..114..127B}. However, the properties of magnetic turbulence in solar flares remain severely lacking in observational constraints. Recent observations of nonthermal spectral line broadening \citep{2021ApJ...923...40S} have revealed widespread turbulence within the flare loop, with the most intense turbulence concentrated at the looptop. Latest 3D MHD simulations \citep[e.g.,][]{2023ApJ...947...67R} have also investigated turbulence in the flare context, particularly in the current sheet and looptop regions, and  obtained consistent results.
Taking the turbulent kinetic energy density as comparable to magnetic fluctuation energy density, we can evaluate the turbulent field $\delta B/B \sim v_{nth}/v_A$ \citep{2017PhRvL.118o5101K}, where $v_{nth}$ is the nonthermal velocity due to turbulence.
In the corona, the Alfvén speed is approximately 1000 km s$^{-1}$, with $v_{nth}$ varying from 10 to 200 km s$^{-1}$ \citep{2021ApJ...923...40S, 2023ApJ...947...67R}, yielding $\delta B/B$ of 1\%$-$20\%. This suggests a relatively weak turbulent field within the flare loop, consistent with the QLT assumption. More robust constraints on flare turbulence from observations and numerical simulations are expected in the future to improve the turbulence model adopted in particle transport simulations.

In the simulations, particle transport processes are primarily governed by the 3D magnetic field and plasma density distributions obtained from the MHD model. Our results highlight the potential of combining data-driven or data-constrained MHD models with particle simulations to understand energy deposition and multi-wavelength emissions during solar flares.
The MHD model is driven by a series of HMI magnetograms. Because of the limited grid resolution, it cannot resolve the plasmoids or mini flux ropes identified in high-resolution 3D MHD simulations under idealized magnetic configurations \citep[e.g.,][]{2017A&A...604L...7M, 2023ApJ...954L..36W, 2025ApJ...992...81W}.
Plasmoids and turbulent magnetic structures in the current sheet and looptop regions may be linked to fine-scale features in flare ribbons and HXR sources \citep[e.g.,][]{2025ApJ...993...31D, 2025ApJ...995L..54F, 2025ApJ...982L...6S}, as well as temporal variability such as quasi-periodic pulsations in flare emissions \citep[e.g.,][]{2022NatCo..13.7680K, 2024ApJ...968....5W, 2026NatAs..10...54A}.  
In addition, recent joint observations from ASO-S/HXI \citep{2024SoPh..299..153S} and Solar Orbiter/STIX \citep{2020A&A...642A..15K} allow for stereoscopic imaging of the HXR emisson \citep[e.g.,][]{2024SoPh..299..114R,2026ApJ...998L..28M}, which will place stringent constraints on models for electron acceleration and transport \citep{2024ApJ...964..145J,2025SoPh..300...56L}.
In future work, we will employ higher-resolution and time-evolving MHD backgrounds, incorporate self-consistent particle acceleration, and generate synthetic HXR and microwave emissions \citep{2024ApJ...971...85C}, to facilitate comprehensive comparisons with multi-wavelength solar flare observations.

\begin{acknowledgments}
X.K. is supported by the National Natural Science Foundation of China (NNSFC) under grants 42574218 and 42561160095, the National Key R\&D Program of China under grants 2022YFF0503002 and 2025YFF0510701, and the Qilu Young Scholars Program of Shandong University.
Z.Z. is supported by the NNSFC under grant 12303061 and the Natural Science Foundation of Shandong Province under grant ZR2023QA074.
G.L. is supported by the Science and Technology Development Fund of Macau (grant No. 002/2024/SKL, 0008/2024/AKP and 0006/2025/RIA1).
The work was carried out at the National Supercomputer Center in Tianjin (TH-3F).
\end{acknowledgments}

\bibliographystyle{aasjournalv7}
\bibliography{export-bibtex}

\end{document}